# Effect of internal chemical pressure on transport properties and TCR for sensors applications

Sudarshan Vadnala[a], Saket Asthana[a*], Prem Pal[a] and S. Srinath[b]

a. Department of Physics, Indian Institute of Technology Hyderabad, ODF Campus, Yeddumailaram, Hyderabad, India-502205.

b. School of Physics, University of Hyderabad, Hyderabad, India.

(*Email: asthanas@iith.ac.in)

**Abstract**

In this work, the structural and transport properties of $(Nd_{0.7-x}La_x)Sr_{0.3}MnO_3$ manganites with x = 0, 0.1 and 0.2 prepared by solid state reaction route are studied. These compounds are found to be crystallized in orthorhombic structural form. The influence of La substitution in place of Nd at A-site shifts the metal to semiconductor/insulator transition temperature ($T_{MI}$) peak towards room temperature with x = 0, 0.1 and 0.2. A composition prepared with the value of x = 0.2 in $(Nd_{0.7-x}La_x)_{0.7}Sr_{0.3}MnO_3$ manganites (i.e. $(Nd_{0.5}La_{0.2})_{0.7}Sr_{0.3}MnO_3$), $T_{MI}$ was observed at 289 K which is close to room temperature. The maximum percentage of TCR values of compounds are increasing with average radius $<r_A>$ but %TCR are slightly equal in x = 0.1 and 0.2 as compared to the parent compound. The maximum %TCR value is almost independent with A-site average radius $<r_A>$ in x = 0.1 and 0.2. The electrical resistivity data are explored by different theoretical models and it has been concluded that at low temperature (ferromagnetic metallic region) conduction mechanism presumably due to the combined effect of electron-electron, electron-phonon and electron-magnon scattering, while in paramagnetic semiconducting regime, the variation of resistivity with temperature are explained by (1) Mott variable range hopping mechanism, (2) Adiabatic small polaron hopping and (3) Thermally activated hopping. The polaron hopping and thermal activation energies are decreasing with increase of an average A-site ionic radius ($<r_A>$). An appropriate enlightenment for the observed behavior is discussed in detail.

# 1. Introduction

In the past few decades, $AMnO_3$ type manganites have been extensively studied because of their richness in physical properties which is due to the simultaneous presence of spin, lattice and orbital degrees of freedom [1-3]. Significant attention has been paid by researchers in order to explore their potential for wide range of technological applications such as read heads, magnetic information storage, low and high field magnetic sensors, IR detectors and more recently spintronic applications [4-11]. The substituted manganites provide high temperature coefficient of resistance (TCR) in bulk as well as in thin films at room temperature that motivates to explore them for infrared radiation detectors (i.e. IR detector) for night vision applications [12]. In perovskites manganite, $NdMnO_3$ is antiferromagnetic insulator characterized by a super exchange coupling between $Mn^{3+}$ sites facilitated by a single $e_g$ electron which is subjected to strong correlation effects. Partial substitution of $Nd^{3+}$ ion with divalent cation results in a mixed valance states of Mn ($Mn^{3+}$ and $Mn^{4+}$) and compound exhibits Metal–Insulator transition ($T_{MI}$) in which metallic regime is of ferromagnetic nature.

The physical properties of the manganites can be tuned either by substituting cations at the A- or B-sites or by varying the oxygen content in the regular perovskite structure [13-15]. The size mismatch at A-site generates internal chemical pressure within the lattice due to the structural distortions of the manganites. The distortion can be controlled by the average size of the A-site cation which in turn modifies the Mn-O-Mn bond angle and Mn-O distances. The Mn-O-Mn bond angle is directly related to the hopping integral between $Mn^{3+}$ and $Mn^{4+}$ degenerate states. Goldschmidt tolerance factor $\tau$ is defined as $\tau = \frac{<r_A>+r_0}{\sqrt{2}(r_B+r_0)}$ where $<r_A>$ and $r_B$ are the radii of the average A-site and B-site ions and $r_0$ is the radius of oxygen ion. For $\tau <1$, Mn-O-Mn bond angle decreases due to rotation of $MnO_6$ octahedra which in turn leads to lower symmetric structure. The transport properties of the manganites are influenced by distortions generated by the size mismatch at A-site due to the cation of different radii of A-site and also by cationic vacancies of rare earth (3+) and divalent alkaline earth (2+) elements which offers a local distortion of the lattice. This disorder is quantified by means of the variance of the A-site cation radius distribution ($\sigma^2$) defined as $\Sigma\ y_i r_i^2 - <r_A>^2$ where $y_i$ are the fractional occupancies of the species. Therefore, the variation of A-site cation radius $<r_A>$ influences the electrical and magnetic properties of the perovskite manganites.

In the present research, we have investigated the structural and transport properties of Nd-based manganites with composition $(Nd_{1-x}La_x)_{0.7}Sr_{0.3}MnO_3$ (where x = 0, 0.1 and 0.2) to tune the TCR and $T_{MI}$ for application aspect. The present work is targeted to achieve favorable properties of manganites for IR detector applications.

## 2. Experimental details

The polycrystalline samples of $(Nd_{0.7-x}La_x)_{0.7}Sr_{0.3}MnO_3$, where x = 0, 0.1, 0.2, are synthesized by the solid state reaction route using ingredients $Nd_2O_3$, $La_2O_3$, $SrCO_3$ and $Mn_2O_3$. The mixed powders are calcined at 1100 °C in air for 24 hours. Thereafter, powder is pressed into pellets by applying a uniaxial pressure of 4-5 tons followed by sintering at 1300 °C for 5 hours. The sintered pellets are annealed in an oxygen environment at 1000 °C for 5 hours to retain the oxygen stoichiometry. The structure and phase purity of the samples are analyzed by powder X-ray diffraction (XRD) performed on a diffractometer (PANanalytic X'pert pro) using Cu $K_\alpha$ radiation at 40 kV and 30 mA. The resistivity measurement without and with magnetic field (5T) are carried out using four-probe method in the temperature range from 5 to 300 K on a quantum design Physical Property Measurement System (PPMS Model No 6000).

## 3. Results & Discussion

*3.1 XRD Results*

The XRD patterns of polycrystalline manganites $(Nd_{0.7-x}La_x)_{0.7}Sr_{0.3}MnO_3$ ( x = 0, 0.1 and 0.2) prepared by solid state route are shown in figure 1. The XRD patterns of all compounds exhibit single–phase orthorhombic unit cell with *Pnma* (No 62, PCPDF Ref No 861534) space group. It can be observed from the patterns that peaks are slightly shifted toward lower angle side on substitution of La in place of Nd. This small shift arises due to the mismatch of radius at A-site i.e., larger radius of La-ion (1.36 Å) in comparison to the radius of Nd-ion (1.27 Å) which causes increase in the volume of the lattice. Subsequently, an internal chemical pressure generated within the lattice due to size mismatch at A-site, which results in slight shift in the peaks of XRD pattern.

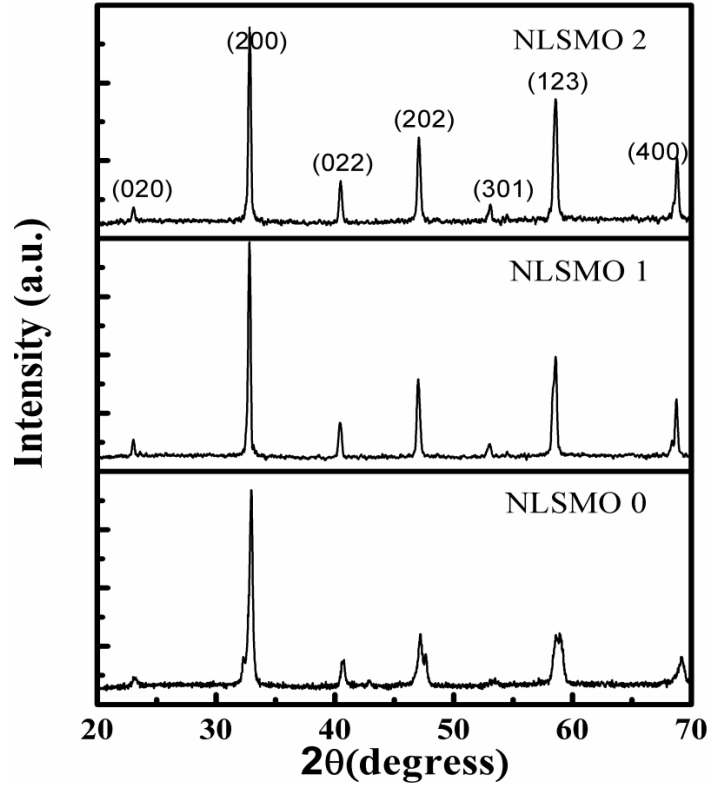

**Figure 1.** XRD patterns of pervoskite manganites $(Nd_{0.7-x}La_x)_{0.7}Sr_{0.3}MnO_3$ where x = 0, 0.1 and 0.2.

*3.2 Electrical transport*

The temperature dependent resistivity ρ(T) studied in the temperature range 5-300 K are shown in figure 2. All samples are exhibiting metal-to-insulator transition ($T_{MI}$). The $T_{MI}$ is shifted from 239 K (for x = 0) to 289 K (for x = 0.2). The average radius of site-A ($<r_A>$) increases as the fraction of La increases. The oxygen ions tends to move towards the center of $MnO_6$ octahedra as $<r_A>$ decreases which in turn leads to distortion. Average A-site ionic radii induced distortion causes reduction in Mn-O bond distances and Mn-O-Mn bond angle. This lattice distortion offers a localized state for the $e_g$ electron and causes possible electronic phase separation within the lattice. Therefore, hopping amplitude of the carriers from $Mn^{3+}$ to $Mn^{4+}$ decreases due to suppression of delocalized hopping sites [16]. As the average radius $<r_A>$ increases, the local lattice distortion reduces due to the shifting of Mn-O-Mn bond angle towards symmetrical side (approaches 180°). As a consequence, hopping amplitude increases which leads to shift in $T_{MI}$ towards higher temperature.

It may be emphasized here that the design and development of uncooled IR detector (Bolometer) require $T_{MI}$ around the room temperature with high TCR for improved sensitivity [12, 17]. In our case, NLSMO 2 is exhibiting $T_{MI}$ close to room temperature and TCR is high as compared to the parent compound NLSMO 0 as shown in the inset of figure 4.

In order to understand the electrical transport mechanism of manganites, the temperature dependent resistivity is categorized into two parts: low temperature (T<$T_{MI}$) and high temperature (T>$T_{MI}$) behaviors. In case of T<$T_{MI}$, TCR is positive (i.e. $\frac{d\rho}{dT}$>0), while it is negative ($\frac{d\rho}{dT}$<0) for T>$T_{MI}$.

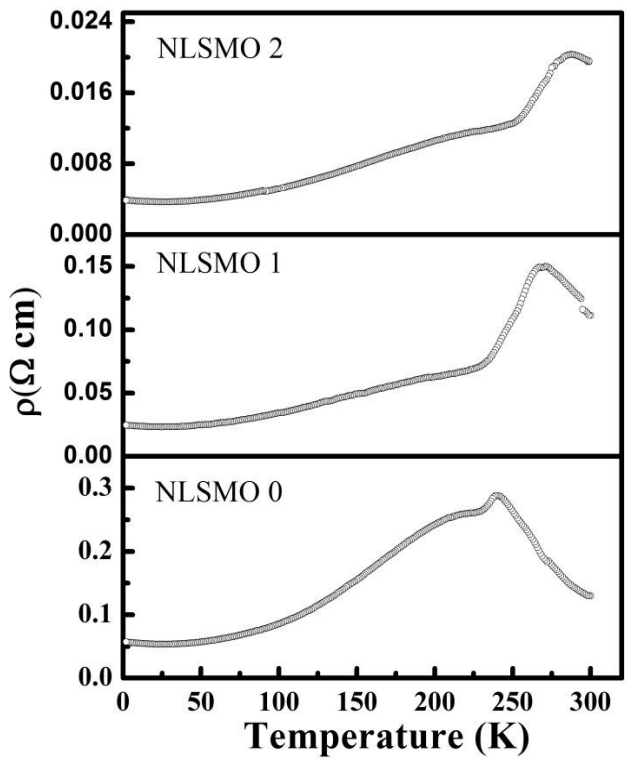

**Figure 2.** Temperature dependent resistivity behavior of $(Nd_{0.7-x}La_x)_{0.7}Sr_{0.3}MnO_3$ where x = 0, 0.1 and 0.2

### 3.2.1 Low temperature (T<$T_{MI}$) behavior

The variation of resistivity at low temperatures (T<$T_{MI}$) and comparative strengths of the different scattering mechanisms originating from several contributions are enlightened on the basis of various equations used to characterize the low temperature resistivity of manganites [18-23]. Here, an attempt has been made to clarify the low temperature resistivity data using the following empirical equations given in Table 1. From these equations, the conduction phenomenon is explained by different scattering mechanisms.

**Table 1.** Various models found in literature to explain the conduction mechanism in low temperature regime (below $T_{MI}$) of Manganites

| Author | Equation | Material | Ref |
|---|---|---|---|
| Schiffer &Vlakhov | $\rho = \rho_0 + \rho_{2.5}5T^{2.5}$ | $La_{0.7}Ca_{0.3}MnO_3$ polycrystalline pellets and film | 18,19 |
| Urushibara | $\rho = \rho_0 + \rho_2 T^2$ | LSMO | 20 |

| Kubo and Ohata | $\rho = \rho_0 + \rho_{4.5}T^{4.5}$ | Doped LaMnO$_3$ | 21 |
| Snyder | $\rho = \rho_0 + \rho_2 T^2 + \rho_{4.5}T^{4.5}$ | La$_{0.67}$A$_{0.33}$MnO$_3$, A=Ca/Sr thin film/bulk | 22 |
| Jaime | $\rho = \rho_0 + \rho_2 T^2 + \rho_5 T^5$ | La$_{0.67}$A$_{0.33}$MnO$_3$, A=Pb/Ca | 23 |

In the equations, $\rho_0$ is the temperature independent residual resistivity which arises due to the grain/domain boundary effects, scattering by impurities, defects and domain walls [24, 25]. As the polycrystalline materials contain grain boundaries, their significant contribution to the resistivity is proved in microwave measurement [26]. Hence, $\rho_0$ plays a major role in the conduction process. The term $\rho_2 T^2$ describes the resistivity due to electron-electron scattering phenomenon [27, 28], while the term $\rho_{2.5}T^{2.5}$ gives resistivity due to the phenomenon of single magnon scattering process in ferromagnetic phase [22, 26, 29]. The last terms $\rho_{4.5}T^{4.5}$ ascribes due to the process of electron-magnon scattering process in the ferromagnetic region [28] and $\rho_5 T^5$ contributes the resistivity due to the phenomenon of electron-phonon interaction [23].

The experimental data of our samples are fitted with general polynomial $\rho = \rho_0 + \rho_2 T^2 + \rho_{4.5}T^{4.5}$ equation and their related fitting graphs and parameters are shown in figure 3 and Table 2 respectively. From these data, the corresponding fitting parameters decrease with increasing the average radius <$r_A$>. It can be observed from the graphs of figure 3, the residual resistivity of the original data is slightly higher than the fitting parameter values. This slight variation in residual resistivity may be due to the presence of grain boundary effects in the polycrystalline materials as well as localization effects particularly at low temperatures. From here it can be concluded that the residual resistivity is decreasing as the average radius <$r_A$> of the composition increasing and vice versa as shown in figure 4 which is consistent with the reported literature [30]. As shown in Table 2, the values of $\rho_2$ and $\rho_{4.5}$ determined from fitting polynomial (figure 3) are also decreasing with increasing value of <$r_A$>.

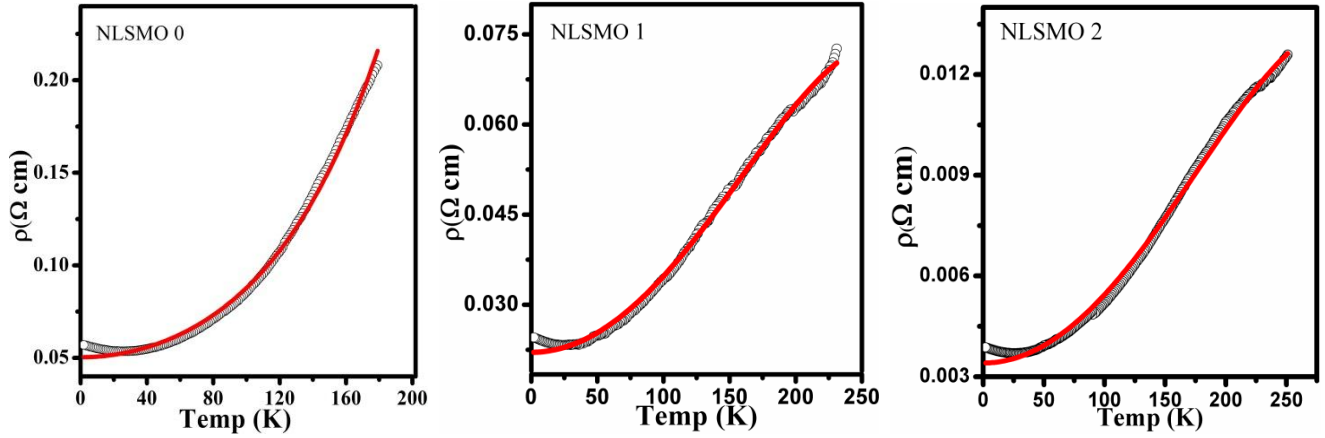

**Figure 3.** Fitted curves of the resistivity data using polynomial equation $\rho = \rho_0 + \rho_2 T^2 + \rho_{4.5} T^{4.5}$ below the metal-insulator transition ($T<T_{MI}$) temperature.

**Table 2.** Theoretical fitting parameters below $T_{MI}$ and maximum %TCR (see figure 4) with respect to average radius $<r_A>$ of NLSMO series.

| Composition | Sample Code | Average radius $<r_A>$ Å | $\rho_0$ ($\Omega$ cm) | $\rho_2$ ($\Omega$ cm K$^{-2}$) | $\rho_{4.5}$ ($\Omega$ cm K$^{-4.5}$) | Maximum TCR% (K$^{-1}$) |
|---|---|---|---|---|---|---|
| $Nd_{0.7}Sr_{0.3}MnO_3$ | NLSMO 0 | 1.321 | 0.05036 | 3.31726E-6 | 4.31377E-12 | 1.4 |
| $Nd_{0.6}La_{0.1}Sr_{0.3}MnO_3$ | NLSMO 1 | 1.330 | 0.02208 | 1.3994E-6 | -5.151E-13 | 2.66 |
| $Nd_{0.5}La_{0.2}Sr_{0.3}MnO_3$ | NLSMO 2 | 1.339 | 0.00341 | 2.1028E-7 | -6.41083E-14 | 2.65 |

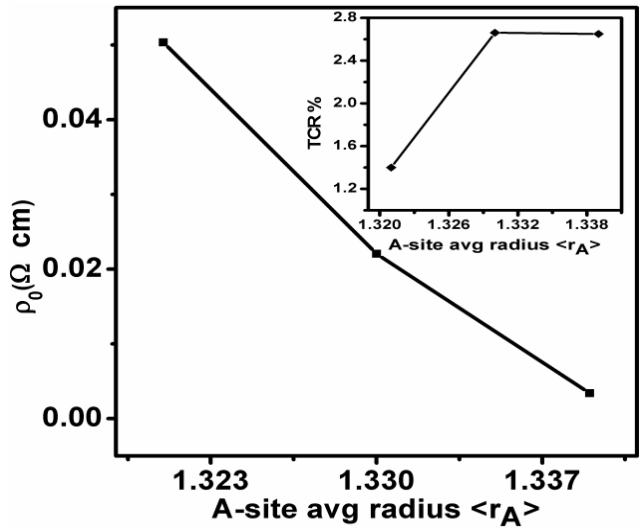

**Figure 4.** Residual resistivity $\rho_0$ vs A-site average radius $<r_A>$. Inset shows maximum TCR% vs A-site average radius $<r_A>$

The observed maximum %TCR $\{= 100 \left[\frac{1}{\rho} * \left(\frac{d\rho}{dT}\right)\right]\}$ vs average radius of A-site <$r_A$> is given in the inset of figure 4. The maximum %TCR values of the compounds are increasing with average radius <$r_A$> but %TCR are slightly equal in x = 0.1 and 0.2 as compared to the parent compound. It is worth to mention here that %TCR has been optimized to a value of 2.66 (for x = 0.1) and 2.65 (for x = 0.2) which are independent with <$r_A$>. This opens the possibility of further improvement in operating temperature for bolometer applications without sacrificing an optimum %TCR value.

### 3.2.2 High Temperature (T >$T_{MI}$) behavior

In paramagnetic or semiconducting/insulating phase, the electrical resistivity generally exhibits strong temperature dependence. Several conduction mechanisms can be used to account the resistivity behavior for T>$T_{MI}$ which are described by three different models viz; Mott variable range hopping model ($T_{MI}$<T< $\theta_D$/2) [31], Small polaron hopping model (T>$\theta_D$/2) where $\theta_D$ is the Debye temperature [32] and the final model is Thermally activated hopping model. On the basis of these models, it is very difficult to say which model is superior to explain the electrical transport properties at high temperature (T>$T_{MI}$). In our case Small polaron hopping model is linearly best fitted as compared with other models. All of the models are linearly fitting very well with good square of linear correlation coefficient ($R^2$) values.

According to Mott Variable Range Hopping model (VRH) the characteristic hopping length increases with lowering temperature and constant density of states are obtained from the well-established Mott's law. Hopping conduction results from the states whose energies are concentrated in a narrow band near the Fermi level is given by the equation $\rho = \rho_0 \exp(T_\circ/T)^{1/4}$ where $T_\circ = 16\alpha^3/k_B N(E_F)$, $k_B$ Boltzmann's constant and $N(E_F)$ is the density of states at the Fermi level. Here we have taken α value 2.22 nm$^{-1}$ which is estimated and reported for manganites in ref. [33]. $(T_\circ)^{1/4}$ values are measured by slope of $\ln(\rho)$ $vs$ $1/T^{1/4}$ is useful for measurement of density of the states at the Fermi level and the values obtained from figure 5 are given in table 3.

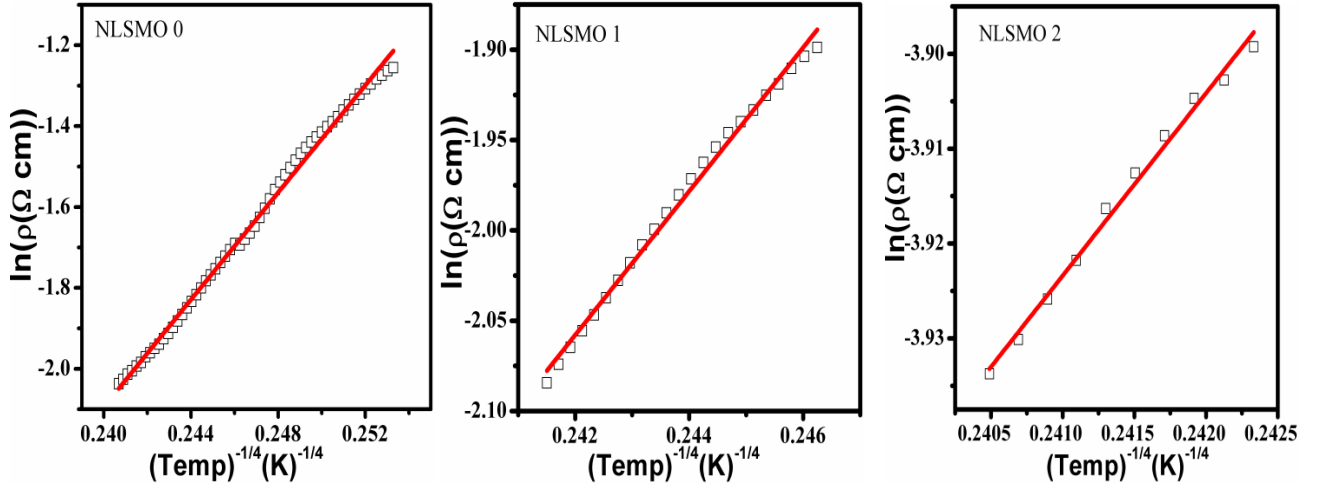

**Figure 5.** Fitted curves of $\ln(\rho)$ $vs$ $1/T^{1/4}$ for $\rho = \rho_0 \exp(T_o/T)^{1/4}$ equation above $T_{MI}$

Small polaron hopping model was used to explain the small polaron influence in the conductivity mechanism is described by polaron models either adiabatic or non-adiabatic equations $\rho = \rho_0 T \exp(E_p/k_B T)$ (adiabatic) and $\rho = \rho_0 T^{3/2} \exp(E_P/k_B T)$ (non-adiabatic) here $\rho_0$ residual resistivity and $E_p$ is the polaron activation energy. Jung et al [34] pointed out that higher value of $N(E_F)$ due to the effect of adiabatic small polaron hopping process. These higher values of the order of ~$10^{22}$ for $N(E_F)$ indicates the clear signatures of the applicability of the adiabatic hopping mechanism. Based on this fact, the adiabatic small polaron hopping model is used in the present investigation rather than non–adiabatic small polaron hopping model. Polaron activation energy ($E_p$) estimated from fitted curves of $\ln(\rho/T)$ vs 1/T is given in table 3. The polaron activation energies are decreasing with increase of average radius $\langle r_A \rangle$ which is due to the Mn-O-Mn bond angle shift towards 180°. Transfer integral defined as $t = \cos(\theta/2)$ is increasing with increase of Mn-O-Mn bond angle, where $\theta$ is an angle between core spins.

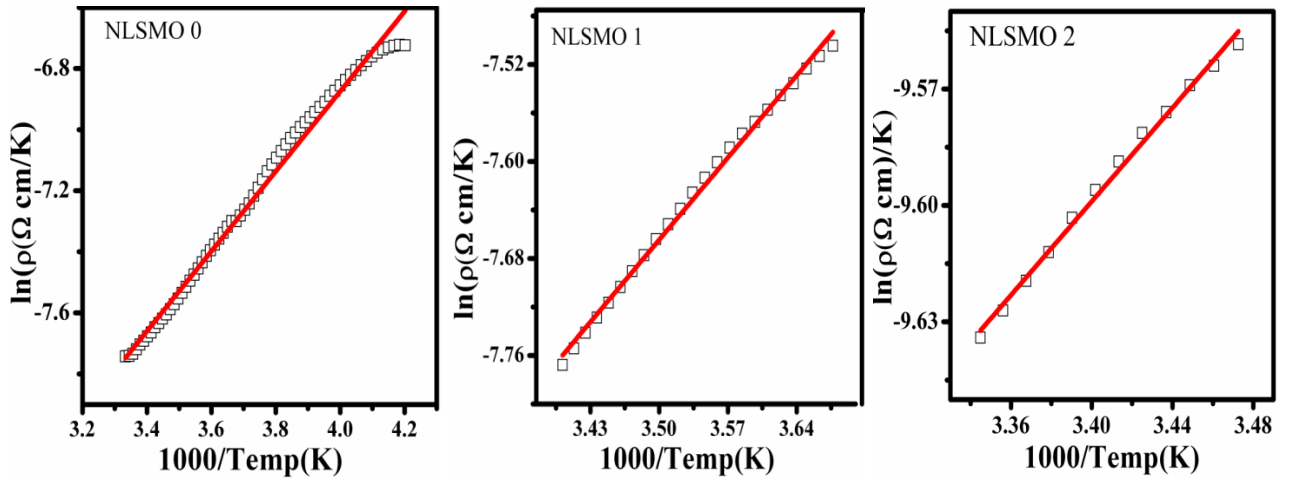

**Figure 6.** Fitted curves of the $\ln(\rho/T)$ $vs$ $1/T$ for adiabatic equation $\rho = \rho_0 T \exp(E_p/k_B T)$ above $T_{MI}$

The thermal activation energy law is given by $\rho = \rho_0 \exp(E_a/k_B T)$ where $E_a$ is the activation energy in semiconducting region. Using this model, obtained activation energy or band gap energy is given in table 3.

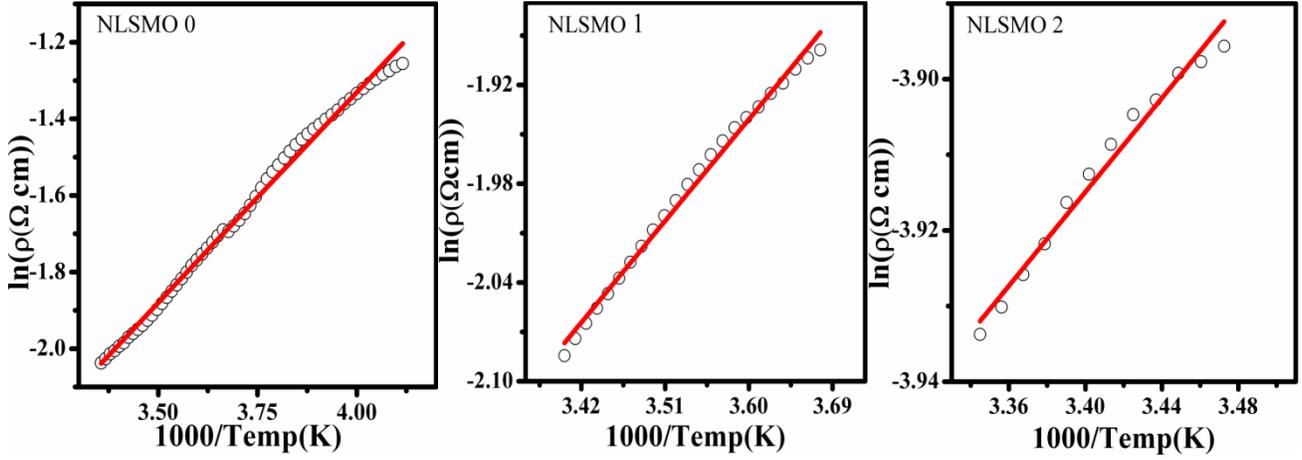

**Figure 7.** Fitted curves of the $\ln(\rho)$ vs $1/T$ for thermal activation energy equation $\rho = \rho_0 \exp(E_a/k_B T)$ above $T_{MI}$

**Table 3.** Fitting parameters by using different models for NLSMO series above $T_{MI}$

| Composition | $(T_o)^{1/4}$ (K$^{1/4}$) | N(E$_F$) (eV$^{-1}$cm$^{-3}$) | E$_p$ (meV) | E$_a$ (meV) |
|---|---|---|---|---|
| NLSMO 0 | 66.2743 | 1.052*10$^{20}$ | 112.9 | 94.8 |
| NLSMO 1 | 39.8132 | 8.083*10$^{20}$ | 83.50 | 59.18 |
| NLSMO 2 | 17.5069 | 2.16*10$^{22}$ | 52.08 | 26.76 |

From table 3 it can be concluded that the polaron activation energy and thermal activation energy are decreasing with increase of the average radius $\langle r_A \rangle$. N(E$_F$) values are increasing with average radius $\langle r_A \rangle$. This N(E$_F$) reflects the carrier effective mass (or narrowing of the band width), which in turn results in a drastic change in the resistivity and sharpening of the resistivity peak at the vicinity of $T_{MI}$ [35]. Higher values of density of states at the Fermi level lead to higher value of conductivity [34].

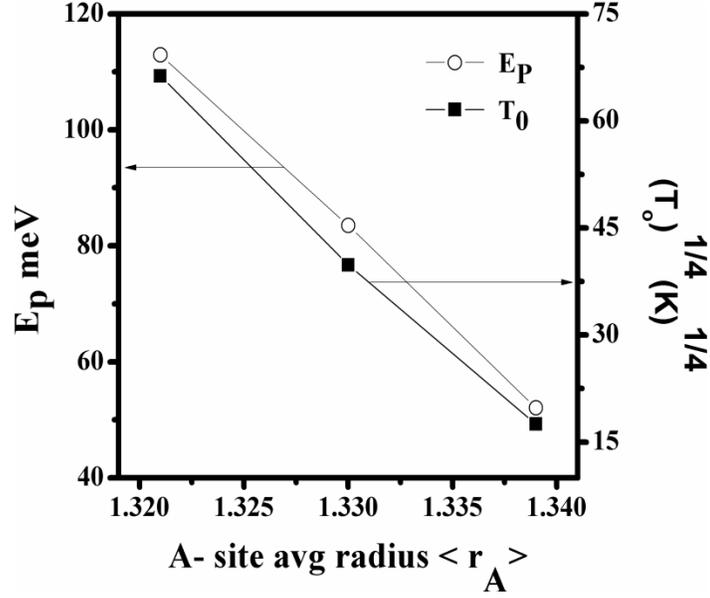

**Figure 8.** Variation of Polaron hopping energy and $(T_o)^{1/4}$ *vs* A-site average radius $\langle r_A \rangle$.

## 3.3 Magnetoresistance

Figure 8 shows the electrical resistivity behavior of the $(Nd_{0.7-x}La_x)_{0.7}Sr_{0.3}MnO_3$ where x = 0, 0.1 and 0.2 samples in a constant magnetic field of 5T. In the presence of an external magnetic field the resistivity decreases significantly. This suggests that the external magnetic field (5T) facilitates the hopping of $e_g$ electron between neighbouring Mn ions, which agrees with the double exchange mechanism [36].

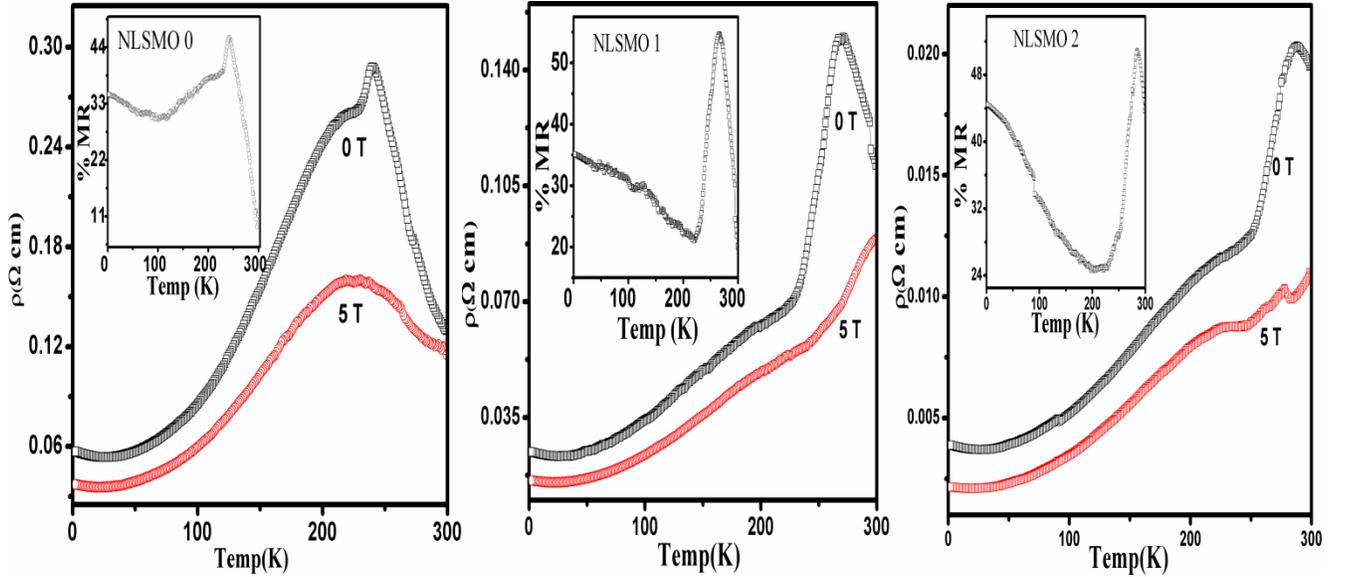

**Figure 9.** Variation of resistivity with temperature of NLSMO series under without and with 5T magnetic field. Inset shows the variation of %MR with temperature.

The magnetoresistance MR is defined as %MR = $100*[\rho(0, T) - \rho(H, T)]/\rho(0, T)$ where $\rho(0, T)$ and $\rho(H, T)$ are the resistivities at temperature without magnetic field and in the applied magnetic field H, respectively. The temperature dependent of the %MR values of the $(Nd_{0.7-x}La_x)_{0.7}Sr_{0.3}MnO_3$ are 46% , 52% and 50% for x = 0, 0.1 and 0.2 as shown in inset of figure 9. The %MR values are increasing slightly higher than the parent compound with varying of the $<r_A>$ this will be useful for spintronic application [8]. The manganese ions are ferromagnetically ordered below $T_{MI}$, therefore within a single magnetic domain the $e_g$ electron transfer between $Mn^{3+}$ and $Mn^{4+}$ ions is easier. The pair of $Mn^{3+}$ and $Mn^{4+}$ spins, which may not be parallel in the vicinity of domain wall boundaries, will act as a hindrance for electron transport. The magnetic domains tend to align in the presence of sufficiently strong magnetic field. As a result, hopping of electrons become easier across the domain wall boundaries and the resistivity decreases, which in turn leads significant MR at low temperature.

## 4. Conclusion

In this paper, the influence of La substitution at Nd site on the electrical and magneto-transport properties in $(Nd_{0.7-x}La_x)_{0.7}Sr_{0.3}MnO_3$ (x = 0, 0.1 and 0.2) has been extensively studied. $T_{MI}$ is shifting towards room temperature with La content. Conduction mechanism behavior is explained by the polynomial equation in low temperature region and the high temperature behavior is described using different existing models. From the present study, it can be concluded that the residual resistivity ($\rho_0$) and the polaron activation energy decreasing with increase of average radius of site-A. An interesting observation is that NLSMO 2 provides $T_{MI}$ around the room temperature and percentage of maximum TCR values are independent with average radius $<r_A>$ in x = 0.1 and x = 0.2. However the TCR value is not satisfactorily high for the development of sensitive microbolometer, this study will be useful in future to optimize the working temperature without affecting the TCR. Moreover, the present research can further be extended for the optimization of the composition to achieve high TCR with $T_{MI}$ around the room temperature for the development of MEMS based uncooled microbolometer for night vision cameras.

**Figure Captions**

Fig.1: XRD patterns of pervoskite manganites $(Nd_{0.7-x}La_x)_{0.7}Sr_{0.3}MnO_3$ where x = 0, 0.1 and 0.2.

Fig.2: Temperature dependent resistivity behavior of $(Nd_{0.7-x}La_x)_{0.7}Sr_{0.3}MnO_3$ where x = 0, 0.1 and 0.2.

Fig.3: Fitted curves of the resistivity data using polynomial equation $\rho = \rho_0 + \rho_2 T^2 + \rho_{4.5} T^{4.5}$ below the metal-insulator transition (T<$T_{MI}$) temperature.

Fig.4: Residual resistivity $\rho_0$ *vs* A-site average radius $<r_A>$. Inset shows maximum TCR% *vs* A-site average radius $<r_A>$

Fig.5: Fitted curves of $\ln(\rho)$ *vs* $1/T^{1/4}$ for $\rho = \rho_0 \exp(T_o/T)^{1/4}$ equation above $T_{MI}$.

Fig.6: Fitted curves of the $\ln(\rho/T)$ *vs* 1/T for adiabatic equation $\rho = \rho_0 T \exp(E_p/K_B T)$ above $T_{MI}$.

Fig.7: Fitted curves of the $\ln(\rho)$ *vs* 1/T for thermal activation energy equation $\rho = \rho_0 \exp(E_a/K_B T)$ above $T_{MI}$.

Fig.8: Variation of Polaron hopping energy and $(T_o)^{1/4}$ *vs* A-site average radius $<r_A>$.

Fig.9: Variation of resistivity with temperature of NLSMO series under without and with 5T magnetic field. Inset shows the variation of %MR with temperature.